\documentclass[prd,twocolumn,showpacs,reprint,preprintnumbers,nofootinbib]{revtex4-2}

\RequirePackage[colorlinks=true
,urlcolor=blue
,anchorcolor=blue
,citecolor=blue
,filecolor=blue
,linkcolor=blue
,menucolor=blue
,linktocpage=true
,pdfproducer=medialab
,pdfa=true
]{hyperref}

\usepackage{amsmath,amssymb,amsthm,amsfonts}
\usepackage{graphicx,tabularx}
\usepackage{color}
\usepackage{multirow}  
\usepackage{comment}
\usepackage{slashed}
\usepackage{enumitem}
\usepackage{cleveref}
\usepackage{rotating}
\usepackage[english]{babel}
\usepackage[justification=centerlast]{caption}
\usepackage{subcaption}
\usepackage{float}

\usepackage{breqn} 
\makeatletter 
\let\cref@old@eq@setnumber\eq@setnumber 
\def\eq@setnumber{% 
\cref@old@eq@setnumber% 
\cref@constructprefix{equation}{\cref@result}% 
\protected@xdef\cref@currentlabel{% 
[equation][\arabic{equation}][\cref@result]\p@equation\theequation}} 
\makeatother

\usepackage{cases}

\crefname{section}{Sec.}{Secs.}
\crefname{figure}{Fig.}{Figs.}
\crefname{equation}{Eq.}{Eqs.}
\crefname{appendix}{Appendix}{Appendices}
\setlist[description]{leftmargin=0.4cm}
\setlist[itemize]{leftmargin=0.4cm}

\newcommand{\be}{\begin{equation}\begin{aligned}}
\newcommand{\ee}{\end{aligned}\end{equation}}

\newcommand{\beq}{\begin{equation}}
\newcommand{\eeq}{\end{equation}}
\newcommand{\beqa}{\begin{eqnarray}}
\newcommand{\eeqa}{\end{eqnarray}}

\newcommand{\mev}{\text{MeV}}
\newcommand{\gev}{\text{GeV}}
\newcommand{\tev}{\text{TeV}}

\newcommand{\m}{\text{m}}

\renewcommand{\eqref}[1]{Eq.~(\ref{#1})}

\newcommand{\eg}{{\em e.g.}}

\newcommand{\TODO}[1]{\textcolor{green}{TODO}}
\RequirePackage[normalem]{ulem}

\DeclareUnicodeCharacter{2212}{\textendash}

\makeatletter
\def\l@subsubsection#1#2{}
\makeatother

\usepackage{multirow}
\usepackage{makecell}

\usepackage{fontawesome}

\begin{document}

%==============================================================================================
\title{Looking forward to photon-coupled long-lived particles II: dark axion portal}

%==============================================================================================

\author{Krzysztof Jod\l{}owski}
\email{k.jodlowski@ibs.re.kr}
% commas problem solution in affiliation from https://tex.stackexchange.com/questions/593986/revtex-problem-with-comma-in-the-addresses?rq=1
\affiliation{Particle Theory and Cosmology Group\char`,{} Center for Theoretical Physics of the Universe\char`,{} Institute for Basic Science (IBS)\char`,{} Daejeon\char`,{} 34126\char`,{} Korea}

\begin{abstract}
The dark axion portal is a dimension-5 coupling between an axion-like particle (ALP), a photon, and a dark photon, which is one of the targets of the intensity frontier searches looking for $\sim\,$sub-GeV long-lived particles (LLPs).
In this work, we re-examine the limits set by existing detectors such as CHARM and NuCal, and by future experiments such as FASER2, MATHUSLA, and SHiP.
We extend previous works by i) considering several mass regimes of the Dark Sector (DS) particles, leading to an extended lifetime regime of the unstable species, ii) including LLPs production occurring in previously neglected vector meson decays that actually dominate the LLP yield, and iii) by implementing secondary LLP production.
It takes place by Primakoff-like upscattering of lighter DS species into LLP on tungsten layers of neutrino emulsion detector FASER$\nu$2.
This process will allow FASER2 to cover a significant portion of the $\gamma c\tau \sim 1\,\m$ region of the parameter space that is otherwise difficult to cover due to the large ($\sim O(100)\,\m$) distance between the primary LLP production point and the decay vessel, where LLP decays take place, which is required in typical beam-dumb experiments for SM background suppression. \href{https://github.com/krzysztofjjodlowski/Looking_forward_to_photon_coupled_LLPs}{\faGithub}
\end{abstract}

\renewcommand{\baselinestretch}{0.85}\normalsize
\maketitle
% \tableofcontents
\renewcommand{\baselinestretch}{1.0}\normalsize

%=============================================================================
\section{\label{sec:intro}Introduction}
%=============================================================================

The dark axion portal (DAP) has recently been proposed \cite{Kaneta:2016wvf,Ejlli:2016asd} as a novel interaction between an axion-like particle (ALP), a dark photon (DP), and photon induced by interactions in the Dark Sector (DS). 
Such a mechanism can take place, \eg, due to 1-loop processes involving massive dark fermions charged under global Peccei-Quinn symmetry $U(1)_{\text{PQ}}$ and gauge groups - $U(1)_{\text{Y}}$ (hypercharge), and $U(1)_{\text{Dark}}$ - which can be viewed as a generalization of the KSVZ \cite{Kim:1979if,Shifman:1979if} axion to DS containing new $U(1)_{\text{Dark}}$ gauge group.

Recent works have shown that the dark axion portal may have interesting astrophysical and cosmological implications that differ from the well-studied photophilic ALP.
DAP can, \eg, provide a new mechanism for the production of dark matter (DM) \cite{Kaneta:2017wfh,Gutierrez:2021gol}, facilitate cosmological relaxation \cite{Choi:2016kke,Domcke:2021yuz}, and lead to axion-photon-dark photon oscillation \cite{Choi:2018dqr,Choi:2018mvk,Choi:2019jwx,Arias:2020tzl,Hook:2021ous}, which affects 21 cm observations, supernovae cooling, and light-shining-through-walls experiments, among other things.
Similarly to an ALP, which is one of the main benchmarks of the searches for $\sim\,$sub-GeV feebly-interacting BSM particles \cite{Battaglieri:2017aum,Beacham:2019nyx,Alimena:2019zri}, the DAP detection prospects have been investigated in B-factories, fixed target neutrino experiments, reactor experiments, and beam dumps \cite{deNiverville:2018hrc,deNiverville:2019xsx,Deniverville:2020rbv}.

In particular, it was shown that the long-lived particle (LLP) displaced decay signature at CHARM, FASER, MATHUSLA and SHiP is particularly effective in covering the $\sim O(100\,\m)$ region of the parameter space.
Since then, the FASER experiment, which began collecting data in 2022 at the start of Run 3 of the LHC, has undergone an intensive research and development phase.
As a result, a dedicated neutrino emulsion detector FASER$\nu$ \cite{FASER:2019dxq,FASER:2020gpr} has been installed in front of the main detector.
Although its main purpose is the detection of collider neutrinos \cite{FASER:2023zcr}, because it is made of tungsten layers, it can also act as a target for the secondary production of LLPs; see \cite{Jodlowski:2019ycu,Jodlowski:2020vhr} for its impact within non-minimal scalar, vector, and sterile neutrino portals.

Another development concerning FASER that motivates our analysis is that, as shown in \cite{Jodlowski:2020vhr}, FASER2 will be sensitive to semi-visible two-body LLP decays.
The final states are an invisible particle (a neutrino or a DS state) and a single high-energy photon: $E_\gamma>0.1\,\tev$ for FASER2\footnote{In fact, recent work \cite{Dienes:2023uve} considered an even lower energy threshold, $E_\gamma>1{\text -}10\,\gev$. We follow the original thresholds from \cite{Jodlowski:2020vhr}, while results for lower energy thresholds, or other changes, can be easily generated using the modified version of $\tt FORESEE$.}, and $E_\gamma>0.1\,\tev$ for FASER$\nu$2.
This opens up a possibility to constrain BSM scenarios with such semi-visible LLP decays in FASER2, where the LLP can be produced in either primary \cite{Dreiner:2022swd,Dienes:2023uve,Kling:2022ehv} or secondary \cite{Jodlowski:2020vhr} production processes.

In this work, we study both of them for DAP, extending the results of \cite{deNiverville:2019xsx} in multiple directions, such as: considering single photon LLP decays at FASER detectors, taking into account secondary LLP production occurring just in front of the main decay vessel, which allows to cover the regime of shorter LLP lifetimes, and electron scattering signatures; we note that in upcoming work we investigate other BSM scenarios using similar signatures \cite{Jodlowski:2023abd,Jodlowski:2023abe}.
Moreover, we study both LLP candidates - when it consists of a dark axion or a dark photon, and our simulation is adapted to the general case.

The paper is organized as follows.
In \cref{sec:model_dark_alp} we discuss the physical aspects of DAP that are relevant to the intensity frontier searches.
In particular, we identify the region of parameter space corresponding to the long-lifetime regime of a dark photon or a dark axion.
In \cref{sec:LLPs_at_exp} we provide the specifics of the LLP production modes and the signatures under consideration, such as: the displaced LLP decays, secondary LLP production, and scattering of DS states with electrons. 
Our main results are discussed in \cref{sec:results}. 
We show sensitivity reach for FASER2, MATHUSLA, NuCal, and SHiP in two mass hierarchies, where either the dark photon or the dark axion act as a LLP.
For both scenarios, we consider several fixed values of the mass ratio between the two DS species, which correspond to different LLP lifetime regimes.
We also compare our results to the case of photophilic ALP.
In \cref{sec:conclusions} we summarize our study.

%=============================================================================
\section{Dark axion portal \label{sec:model_dark_alp}}
%=============================================================================

The interaction Lagrangian of the dark axion portal is \cite{Kaneta:2016wvf,Ejlli:2016asd},
\be
\!\!\mathcal{L} & \supset \frac{g_{a\gamma\gamma^{\prime}}}{4} a F^{\mu\nu} \tilde{F'}_{\mu\nu}\,,
\label{eq:L_dark_axion}
\ee
where $g_{a\gamma\gamma^{\prime}}$ is a coupling of mass-dimension -1, and $F_{\mu\nu}$ and $F'_{\mu\nu}$ are the EM and $U(1)_{\text{Dark}}$ field strength tensors, respectively.

As described in \cref{sec:intro}, the dark axion portal leads to an interesting range of phenomena that can be distinct from the photophilic ALP.
In particular, dark axion portal was proposed \cite{deNiverville:2018hrc} as an explanation of the recently rejuvenated $(g-2)_\mu$ anomaly \cite{Muong-2:2006rrc,Muong-2:2015xgu,Keshavarzi:2019bjn,Muong-2:2021ojo}.
The region of parameter space relevant to such a solution was the long-lived dark photon with $\sim\,$GeV mass.
In fact, \cite{deNiverville:2018hrc} analyzed the dark photon displaced decays in the past beam dump and neutrino experiments: LSND \cite{LSND:1996jxj}, MiniBooNE \cite{MiniBooNE:2017nqe,MiniBooNE:2008paa}, and CHARM \cite{CHARM:1985anb} (missing energy searches at BaBar \cite{BaBar:2001yhh,BaBar:2013agn} and Belle \cite{Belle-II:2010dht} were also considered, and were shown to provide the coverage of the high mass range) to exclude such possibility.
On the other hand, an extended dark axion portal, involving also kinetic mixing with SM hypercharge or muon-philic interactions, has been shown to be a viable solution \cite{Ge:2021cjz,Zhevlakov:2022vio}. 
Moreover, such a scenario could be tested in future lepton fixed target experiments, such as NA64$e$ \cite{Banerjee:2019pds,NA64:2021xzo}, NA64$\mu$ \cite{Sieber:2021fue}, LDMX \cite{Mans:2017vej}, and M$^3$ \cite{Kahn:2018cqs}.
These considerations further motivate dedicated sensitivity study of the long-lifetime regime of the DAP at the far-forward region of the LHC.

In the following, we discuss two benchmarks where, in each case, one of the DS species is massless, stable particle. 
Then only the coupling $g_{a\gamma\gamma^{\prime}}$ and the LLP mass are free parameters of the model.
In \cref{sec:results} we present results for both of these benchmarks, as well as for several additional scenarios in which the masses of the DS states follow a fixed ratio.

For both benchmarks, the lifetime of the unstable, and typically long-lived, particle depends on the width of the two-body decay into a photon and a DS state.
The three-body decays into a pair of charged leptons and a DS state are also possible, especially for $m\gtrsim 0.1\,\gev$, but they are phase-space suppressed. 
As a result, they will contribute to the total decay width typically only at the $O(0.01)$ level - see Fig. 1 in \cite{deNiverville:2019xsx}; relevant formulas are given in \cref{app:three_body_decays}.

Since FASER detectors are $\sim 400-600\,\m$ away from the $p{\text-}p$ collision point of the LHC, the typical LLP decay lengths they can probe are 
\be
  d_{\gamma^\prime} \simeq &\, 100 \,\m \times \left(\frac{E}{1000\,\gev}\right) \left(\frac{0.1\,\gev}{m_{\gamma^\prime}}\right)^4 \left(\frac{7 \times 10^{-5}}{g_{a\gamma\gamma^\prime}}\right)^2,
  \label{eq:ctau_gprime}
\ee
for the massless dark axion, while for the massless dark photon analogous formula for $d_a$ holds for $g_{a\gamma\gamma^\prime}=4 \times 10^{-5}\,\gev^{-1}$.
We used $d = c \beta \tau \gamma$, where $\gamma = E/m$ is the boost factor of LLP in the LAB frame, $\beta = \sqrt{1-1/\gamma^2}$, and $\tau = 1 / \Gamma$.
The decay widths for the two-body final states are \cite{Kaneta:2016wvf}
\begin{dmath}[labelprefix={eq:}]
  {\Gamma_{\gamma^{\prime} \to \gamma a} = \frac{g_{a\gamma\gamma^{\prime}}^2}{96 \pi} m_{\gamma^{\prime}}^3 \left(1-\frac{m^2_a}{m^2_{\gamma^{\prime}}}\right)^3,} \\
  {\Gamma_{a\to \gamma \gamma^{\prime}} = \frac{g_{a\gamma\gamma^{\prime}}^2}{32 \pi} m_a^3 \left(1-\frac{m^2_{\gamma^{\prime}}}{m^2_a}\right)^3.}
  \label{eq:Gamma_two_body}
\end{dmath}
We note that when one DS particle is massless, the lifetime of $a$ is smaller than the lifetime of $\gamma^\prime$ by a factor of $3$, coming from the average over dark photon polarization states. The same factor will occur for other pairs of processes in which $a$ and $\gamma^{\prime}$ are exchanged which will influence our results in \cref{sec:results}.

\begin{figure}[tb]
  \centering
  \includegraphics[width=0.48\textwidth]{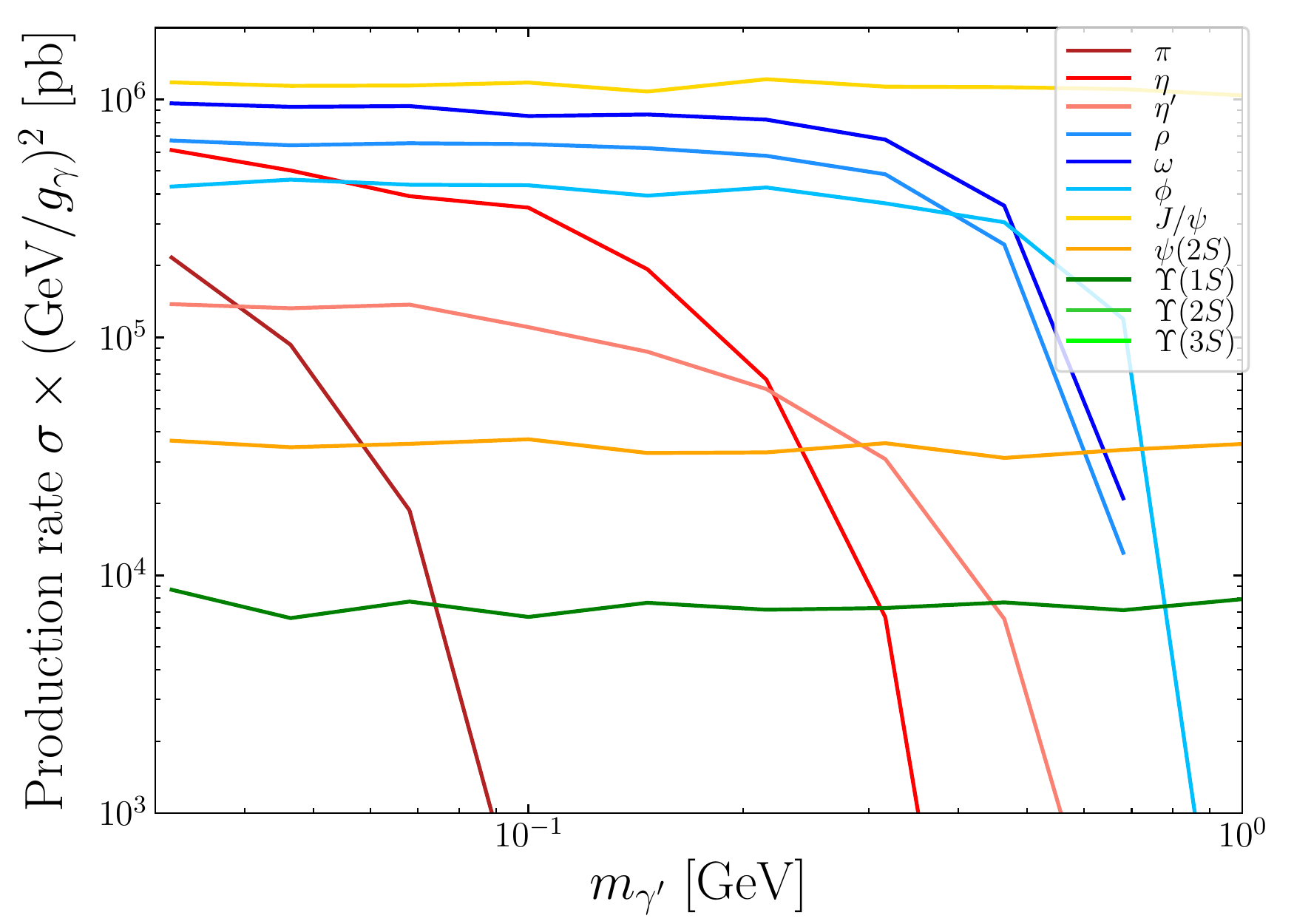}
  \caption{Dark photon production modes as a function of its mass. The vector meson decays, which were not included in previous works, dominate at both FASER2 and SHiP.
  The same relationship exists for other mass schemes and for the dark axion production yields.
  }
  \label{fig:modes_production}
\end{figure}

\begin{table*}[tb]
  \centering
  \begin{tabular}{|c||c|c|c|c|c|c|c|c|c|}
    \hline
    \hline
    Experiment & \thead{Target for\\ prim./sec. \\ prod.} & Energy & \thead{Lumi.\\or $N_{\mathrm{prot.}}$} & \thead{Transverse \\ size} & $x_{\mathrm{min}}$ & $\Delta$ & LLP signature & \thead{LLP \\signature cuts} & Ref. \\
    \hline
    \hline
    CHARM & Cu/- & 400 GeV & $2.4 \times 10^{18}$ & $3\times 3$ m$^2$\footnote{CHARM was placed at a distance of 5 m from the beam axis.} & 480 m & 35 m & decay & \thead{$E_{e^+ e^-}>3\ \gev$: $N_{\mathrm{ev}}=3$ \\  $E_{\gamma}>7.5\ \gev$: $N_{\mathrm{ev}}=100$} & \cite{deNiverville:2018hrc,Dobrich:2019dxc} \\
    \hline
    NuCal & Fe/- & 69 GeV & $1.7 \times 10^{18}$ & $r=1.3$ m & 23 m & 64 m & decay & \thead{$E_{e^+ e^-}>10\ \gev$: $N_{\mathrm{ev}}=4.4$ \\ $E_{\gamma}>10\ \gev$: $N_{\mathrm{ev}}=4.4$} & \cite{Blumlein:2013cua,Dobrich:2019dxc} \\
    \hline
    SHiP & Mo/- & 400 GeV & $2.4 \times 10^{18}$ & $2.5\times 5.5$ m$^2$ & 52.7 m & 50 m & decay & \thead{$E_{e^+ e^-}>3\ \gev$: $N_{\mathrm{ev}}=3$ \\  $E_{\gamma}>2\ \gev$: $N_{\mathrm{ev}}=100$}  & \cite{deNiverville:2018hrc,Jodlowski:2019ycu,Dienes:2023uve} \\
    \hline
    \hline
    FASER2 & Fe\footnote{By primary LLP production at the LHC, we mean the Primakoff process in which photons produced in pp collisions hit the iron hadronic absorber TAN located 140 m further converting into a LLP particle; the same is assumed for other vesions of FASER detector. 
    This production mode has been used for photophilic ALP \cite{Feng:2018pew} and massive spin-2 portal \cite{Jodlowski:2023yne}.}/- & \thead{$\sqrt{s}=$ \\ $13\,\tev$} & $3000$ fb$^{-1}$ & $r=1$ m & 480 m & 5 m & decay  & \thead{$E_{e^+ e^-}>100\ \gev$: $N_{\mathrm{ev}}=3$ \\  $E_{\gamma}>100\ \gev$: $N_{\mathrm{ev}}=3$}  & \cite{Feng:2018pew,Jodlowski:2020vhr} \\
    \hline
    FASER$\nu$2 & Fe/W & \thead{$\sqrt{s}=$ \\ $13\,\tev$} & $3000$ fb$^{-1}$ & $r=0.25$ m & 472 m & 2 m & \thead{decay, \\ sec. prod., \\ $e^-$ scat.} & \thead{$E_{\gamma}>1000\ \gev$: $N_{\mathrm{ev}}=3$, \\ $300\, \mev<E_{e^-}< 20\, \gev$:\footnote{For FASER$\nu$2 and FLArE we also take into the account the angular cuts - see tables 1 and 2 from \cite{Batell:2021blf}.} \\ $N_{\text{ev}}=20$} & \cite{Jodlowski:2019ycu,Jodlowski:2020vhr} \\
    \hline
    FPF FASER2 & Fe/- & \thead{$\sqrt{s}=$ \\ $13\,\tev$} & $3000$ fb$^{-1}$ & $r=1$ m & 620 m & 25 m & decay  & \thead{$E_{e^+ e^-}>100\ \gev$: $N_{\mathrm{ev}}=3$ \\  $E_{\gamma}>100\ \gev$: $N_{\mathrm{ev}}=3$} & \cite{Feng:2018pew,Feng:2022inv,Jodlowski:2020vhr} \\
    \hline
    FPF FASER$\nu$2 & Fe/W & \thead{$\sqrt{s}=$ \\ $13\,\tev$} & $3000$ fb$^{-1}$ & $0.4\times 0.4$ m$^2$ & 612 m & 8 m & \thead{decay, \\ sec. prod., \\ $e^-$ scat.}  & \thead{$E_{\gamma}>1000\ \gev$: $N_{\mathrm{ev}}=3$, \\ $300\, \mev<E_{e^-}< 20\, \gev$: \\ $N_{\text{ev}}=20$} & \cite{Jodlowski:2019ycu,Feng:2022inv,Jodlowski:2020vhr} \\
    \hline
    FPF FLArE &  Fe/Ar & \thead{$\sqrt{s}=$ \\ $13\,\tev$} & $3000$ fb$^{-1}$ & $1\times 1$ m$^2$ & 600 m &  7 m & \thead{sec. prod.,  \\ $e^-$ scat.}  & \thead{$30\, \mev<E_{e^-}< 1\, \gev$: \\ $N_{\mathrm{ev}}=20$} & \cite{Batell:2021blf,Kling:2022ykt,Kling:2022ehv,Feng:2022inv} \\
    \hline
    \hline
  \end{tabular}
  \caption{
    Technical parameters of the considered detectors sensitive to LLP decays, secondary LLP production or scattering with electrons. 
    We specify the technical parameters with the references of each experiment used in our simulations.
    LHC-based detectors are separated from experiments using dedicated proton beams. See Tab. 1 from \cite{Jodlowski:2023yne} for a table of experiments sensitive to a massive spin-2 particle decaying dominantly into a photon pair.
  }
  \label{tab:experiments}
\end{table*}

\begin{figure*}[tb]
  \centering
  \includegraphics[width=0.44\textwidth]{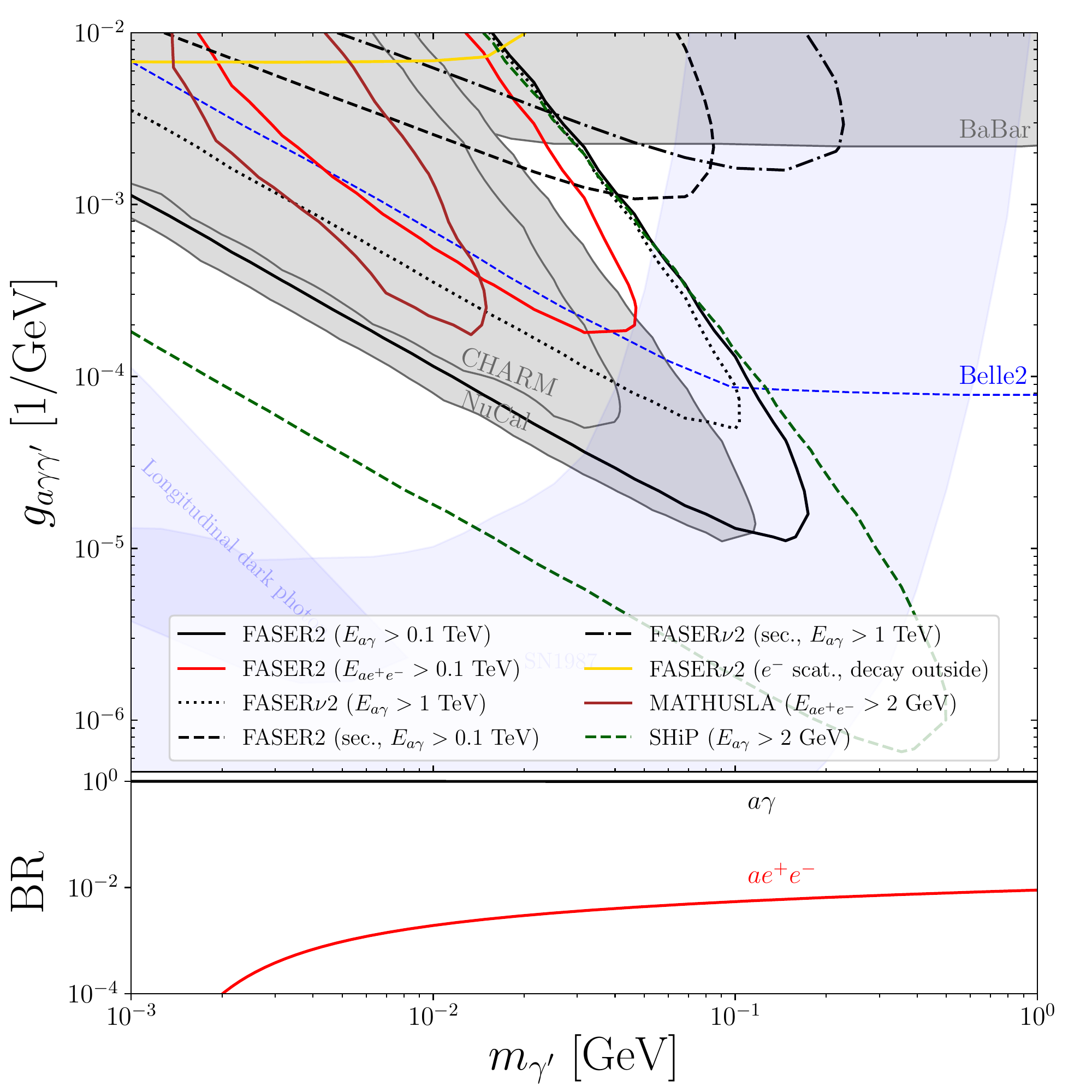}
  \hspace*{0.4cm}
  \includegraphics[width=0.44\textwidth]{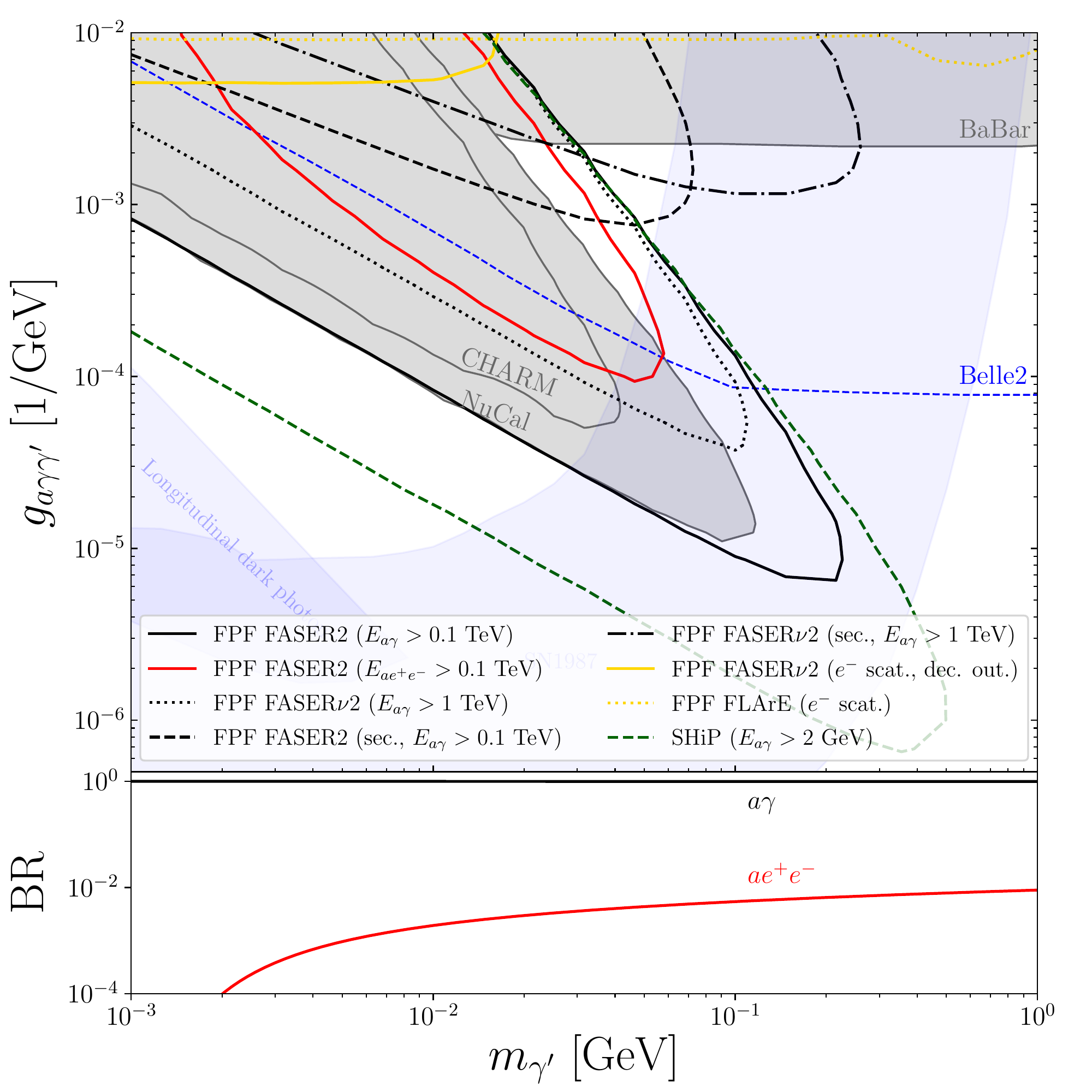}
  % \vspace*{0.2cm}
  \includegraphics[width=0.44\textwidth]{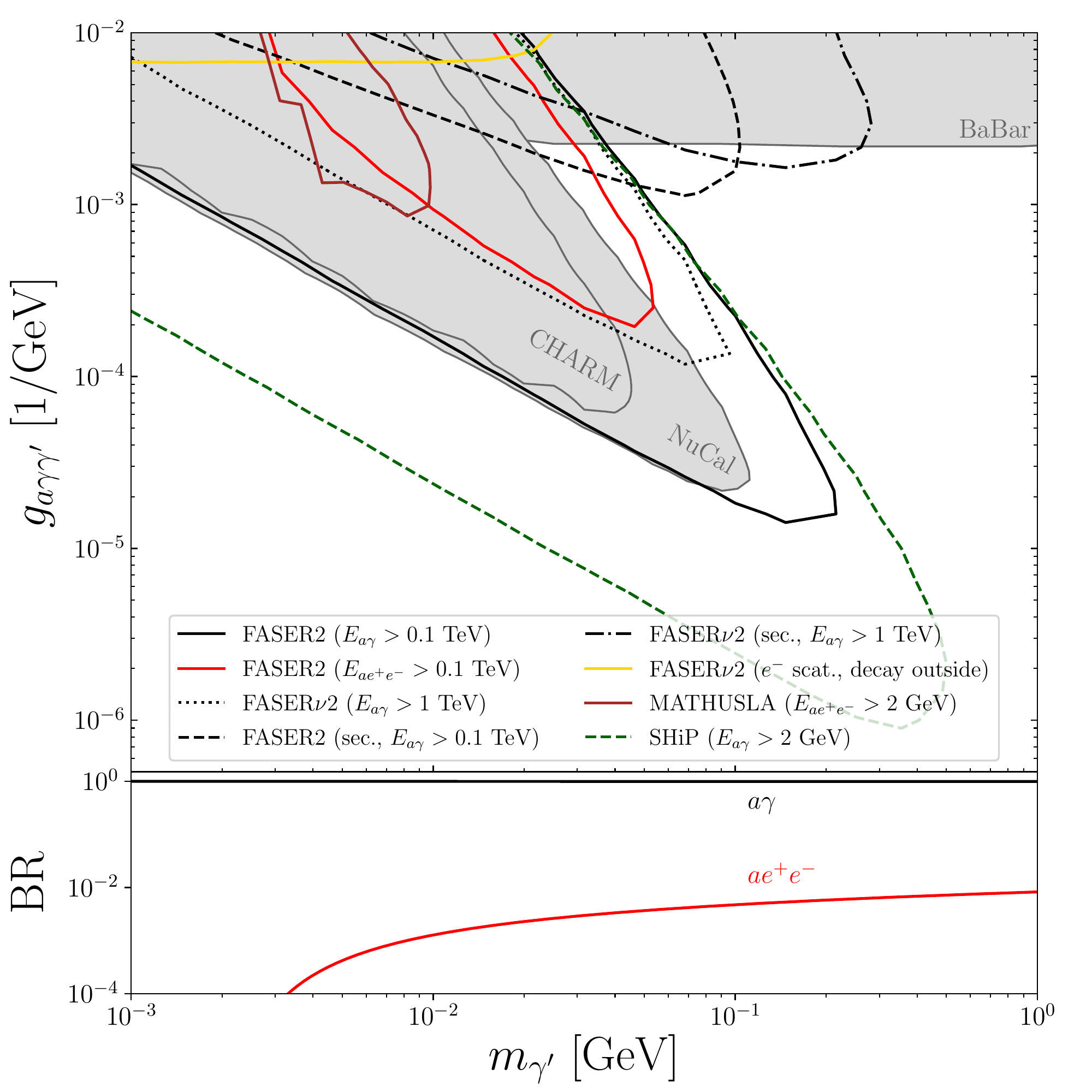}
  \hspace*{0.4cm}
  \includegraphics[width=0.44\textwidth]{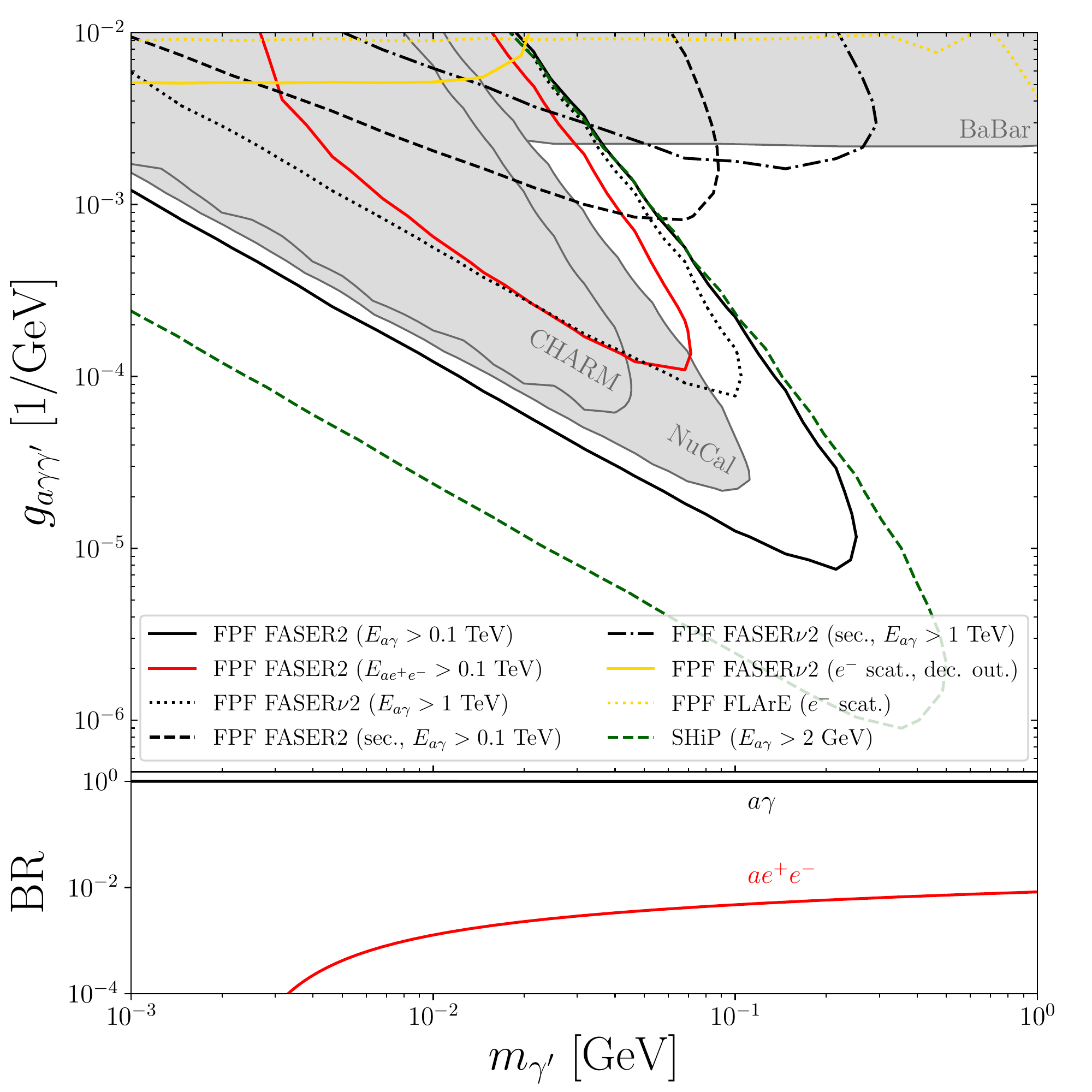}
  \includegraphics[width=0.44\textwidth]{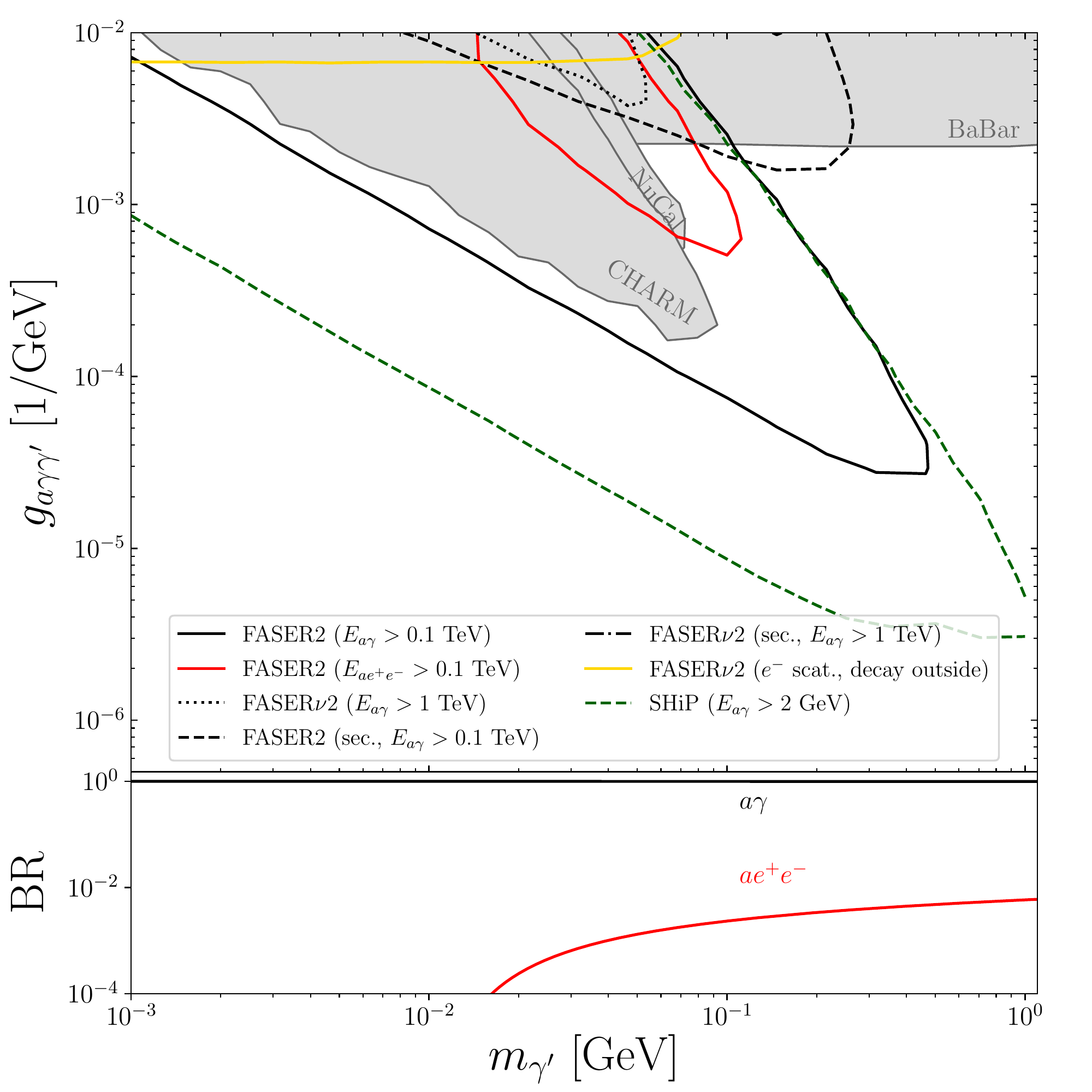}
  \hspace*{0.4cm}
  \includegraphics[width=0.44\textwidth]{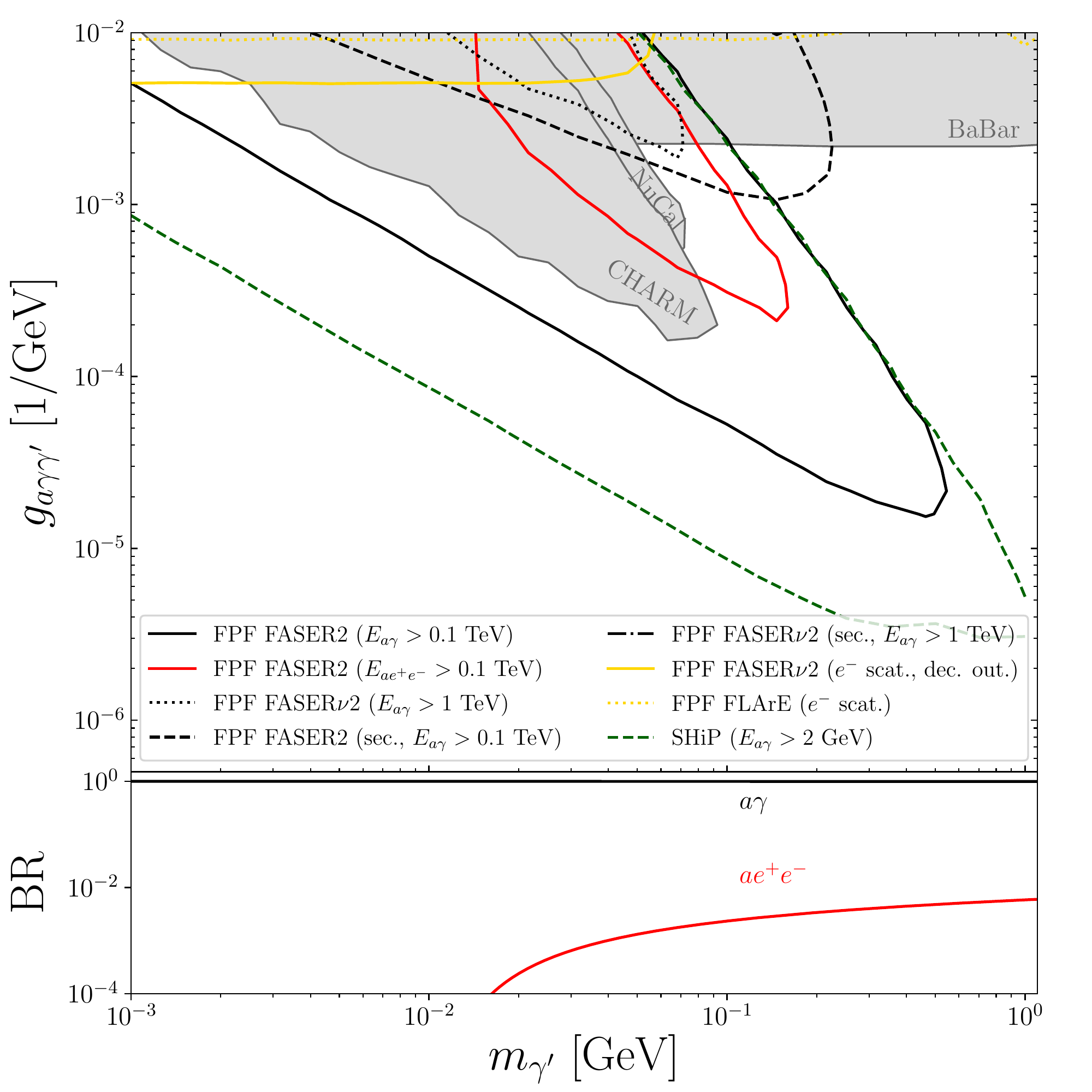}
  \caption{
    Sensitivity reach for the dark photon acting as the LLP at the baseline (left) and the Forward Physics Facility (right) location of FASER2.
    The mass ratio, $m_{a}/m_{\gamma^\prime}$, is fixed as follows: $0$ (top), $0.5$ (middle), and $0.9$ (bottom).
    The contour lines for each experiment correspond to the number of events, $N_{\mathrm{ev}}$, as indicated in \cref{tab:experiments}.
    Lines derived by the missing energy signature at BaBar and Belle were taken from \cite{deNiverville:2018hrc}.
    }
  \label{fig:results_dark_photon}
\end{figure*}

\begin{figure*}[tb]
  \centering
  \includegraphics[width=0.48\textwidth]{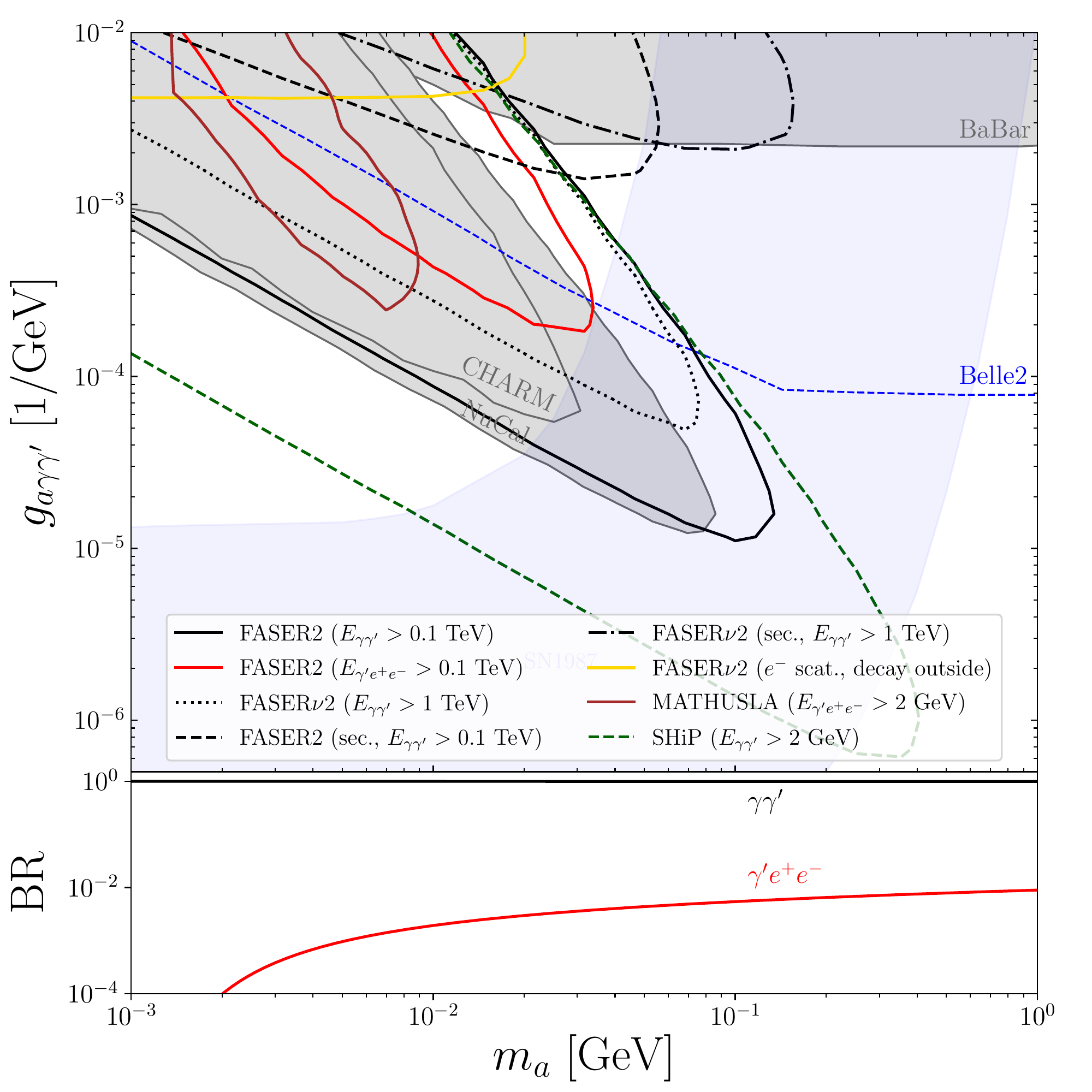}\hspace*{0.4cm}
  \includegraphics[width=0.48\textwidth]{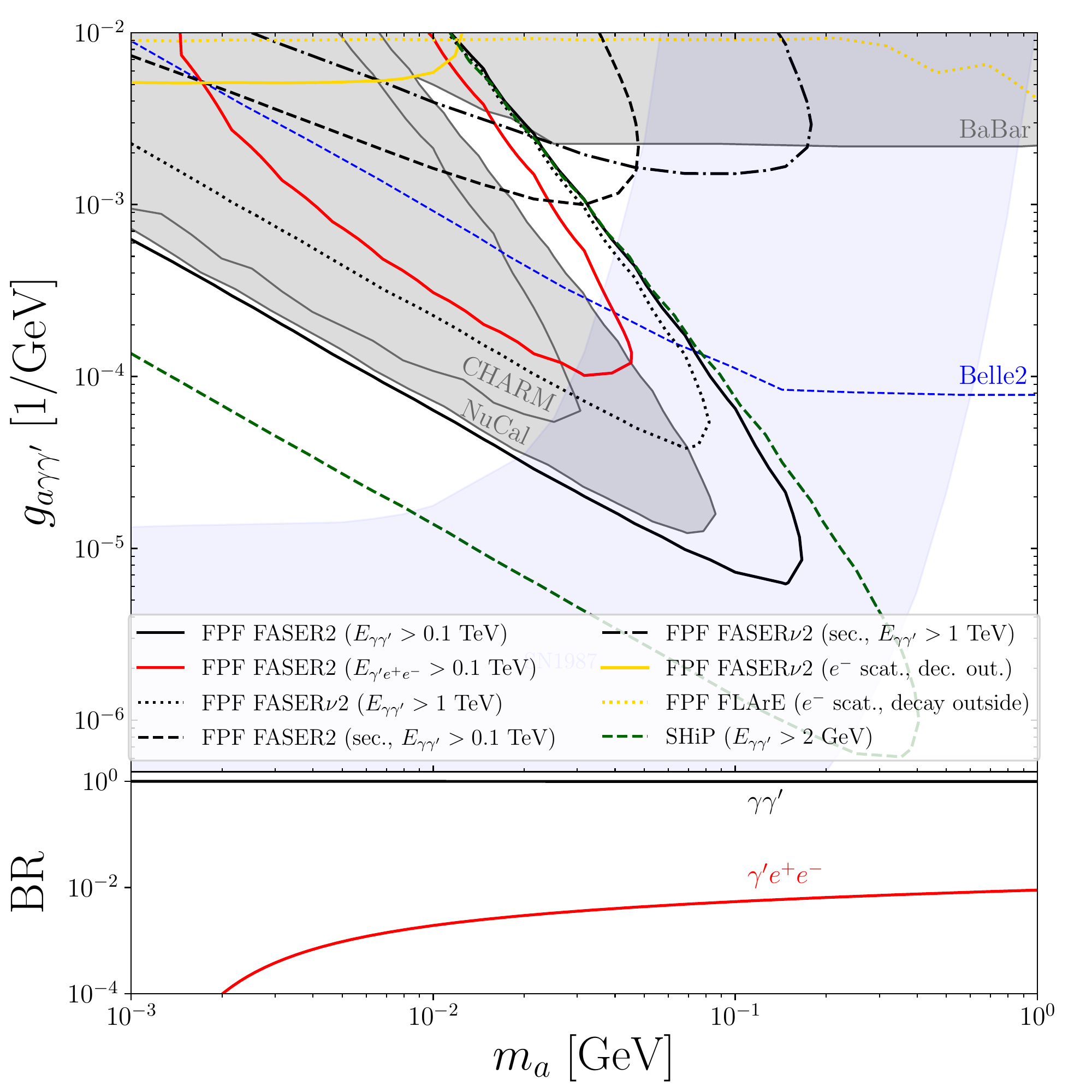}
  \caption{
    Same as \cref{fig:results_dark_photon} but for dark axion acting as the LLP. We only show results for one mass scheme, massless dark photon, as the other results are analogous to the middle and bottom plots of \cref{fig:results_dark_photon}.
    The light-gray areas are excluded by astrophysical and cosmological bounds which were obtained in \cite{Hook:2021ous}.
  }
  \label{fig:results_dark_axion}
\end{figure*}

%=============================================================================
\section{LLP signatures\label{sec:LLPs_at_exp}}
%=============================================================================

In this section, we describe the LLP signatures we use to constrain the DAP, followed by details of the beam dump and LHC experiments under consideration.

%==========================================================================
\subsection{LLP production\label{sec:llp_prod}}
%==========================================================================

In the ${\sim} 1\,\mev {\text-} 1\,\gev$ mass range, dark axion and dark photon are mainly produced from decays of unstable mesons. 
Compared to photophilic ALP, for which the Primakoff conversion of an on-shell photon into ALP dominates, the result is about an order of magnitude smaller number of LLPs for DAP.

Moreover, with regard to meson decays, previous works \cite{deNiverville:2018hrc,deNiverville:2019xsx} considered only three-body decays of pseudoscalar mesons.
% , while, 
As discusssed in \cite{Chu:2020ysb,Dienes:2023uve} for dark fermions coupled to the SM via dimension-5 and dimension-6 electromagnetic form factors, the branching ratios of decays of pseudoscalar and vector mesons of mass $M$ into such DS states are approximately proportional to $M^2$.
As a result, the heaviest vector meson produced in sufficiently large quantities dominate the pseudoscalar mesons contributions.
We found agreement with this argument for the DAP, as shown in \cref{fig:modes_production}, where we present contributions of unstable mesons produced at the LHC to the dark photon yield assuming the dark axion is massless; analogous behavior takes place for other mass scenarios.

We used the $\tt FORESEE$ \cite{Kling:2021fwx} package to implement our model, in particular we used \cref{eq:br_vec} and \cref{eq:br2dq2dcostheta} describing the branching ratios of vector and pseudoscalar meson decays, respectively, which are used to obtain the resulting LLP yield.
For the far-forward LHC detectors such as FASER2 \cite{FASER:2018ceo,FASER:2018bac,FASER:2021ljd} and Forward Physics Facility (FPF) \cite{MammenAbraham:2020hex,Anchordoqui:2021ghd,Feng:2022inv}, which would accommodate multiple detectors adapted  to various searches, \eg, FASER$\nu$2 \cite{Batell:2021blf,Anchordoqui:2021ghd} and FLArE \cite{Batell:2021blf}, we used the included spectra generated by $\tt EPOSLHC$ \cite{Pierog:2013ria} and $\tt Pythia$ \cite{Sjostrand:2014zea}.
For beam dump experiments such as CHARM \cite{CHARM:1985anb}, NuCal \cite{Blumlein:1990ay,Blumlein:2011mv}, and SHiP \cite{SHiP:2015vad,Alekhin:2015byh}, we used $\tt Pythia$ to generate the meson spectra, and we extended $\tt FORESEE$ to simulate the production and decay of LLPs taking place in these detectors.

%==========================================================================
\subsection{Simulation details\label{sec:simulation}}
%==========================================================================

After the production of a LLP, the number of events linked to a LLP signature being detected inside the detector are \cite{Bauer:2018onh,Feng:2017uoz}
\be
  N = \int \int dE d\theta \frac{d^2 N}{dE d\theta}\, p(E, \theta)\, q_{\text{accept.}}(E, \theta),
  \label{eq:NoE}
\ee
where the first term denotes the spectrum of the LLP with a energy $E$ and polar angle $\theta$ relative to the beamline; $p(E)$ corresponds to the probability of the signature taking place inside the detector, while experimental or simulation-related cuts are described by $q_{\text{accept.}}(E, \theta, \phi)$.

\paragraph{Primary production}

Displaced LLP decays resulting from, \eg, proton-target collisions are the main experimental signature in LLP searches \cite{Battaglieri:2017aum,Beacham:2019nyx,Krnjaic:2022ozp}.
The experimental signal consists of high-energy SM particles, typically a pair of photons or charged leptons, and the probability of these decays occurring within a detector of length $\Delta$ is
\be
  p(E) = e^{-L/d(E)}-e^{-(L+\Delta)/d(E)},
  \label{eq:p_prim}
\ee
where $d(E)$ represents the LLP decay length in the LAB frame and $L$ corresponds to the distance between the LLP production point and the start of the detector. 
It is evident that the majority of events arise from sufficiently long-lived species, characterized by $d \gtrsim L$, resulting only in linear suppression with the decay length: $p(E) \simeq \Delta/d$ \cite{Essig:2013lka,Beacham:2019nyx}.
However, for short-lived species, the second term in \cref{eq:p_prim} can be neglected and $p(E) \simeq e^{-L/d}$.
It is therefore clear that the distance $L$ sets the scale of the LLP decay lengths that can be probed in such a way.

In the DAP, the leading two-body decays deposit energy through a single photon, while decays into a DS and $e^+ e^-$ are suppressed, see bottom panels of each plot in \cref{fig:results_dark_photon,fig:results_dark_axion}.
Despite the additional SM induced background for the single-photon LLP decay, it was shown \cite{Jodlowski:2020vhr} that FASER2 will be sensitive to it with the same cuts on the deposited energy and number of events as for the two-photon decays; we refer to that work for discussion of the backgrounds.

\paragraph{Secondary production}

Secondary production of LLPs can take place by coherent upscattering of a lighter DS species into the LLP on tungsten layers of neutrino emulsion detector FASER$\nu$2; see fig. 1 from \cite{Jodlowski:2019ycu} for a schematic illustration.

We study the displaced decay of the LLP produced in this way, where the production takes place at FASER$\nu$2, while the decay happens at FASER2.
As the distance between these two detectors is $L \simeq 1\,\m$, this production mode could allow to cover a part of the $d \sim L \simeq 1\,\m$ region of the parameter space.
On the other hand, the cross-section for the secondary production results in additional $\propto g_{a\gamma\gamma^{\prime}}^2$ dependence in the number of decays. 
As a result, we expect secondary production to cover larger values of $g_{a\gamma\gamma^{\prime}}$ than the ones covered by primary production \cite{Jodlowski:2019ycu,Jodlowski:2020vhr}.

The probability of secondary LLP production followed by decay inside FASER2 is given by convolution of \cref{eq:p_prim} with upscattering cross-section \cite{Jodlowski:2019ycu}
\begin{widetext}
  \be
    p(E)_{\text{sec.\! prod.}} = \frac{1}{L_{\text{int}}} \int_{0}^{\tilde{\Delta}} \left(e^{-(x_{\text{min}}-t)/d}-e^{-(x_{\text{min}}+\Delta-t)/d}\right)\, dt = \frac{d}{m_T/(\rho\, \sigma(E))} e^{-(x_{\text{min}}+\Delta)/d} \left(e^{\Delta/d}-1\right) \left(e^{\tilde{\Delta}/d}-1\right),
    \label{eq:p_sec}
  \ee
\end{widetext}
where $L_{\text{int}}=m_T/(\rho\, \sigma(E))$ is the interaction length corresponding to the upscattering of DS species with energy $E$ on nucleus of mass $m_T$ inside the material of density $\rho$ and length $\tilde{\Delta}$; $\sigma(E)$ is the upscattering cross-section; $x_{\text{min}}$ is the distance from the beginning of the upscattering material to the beginning of the detector; and the dummy variable $t$ parameterizes the length of the upscattering material.

The cross-section for the upscattering process can be obtained in the closed form following the method described for photophilic ALP \cite{Dusaev:2020gxi}; also see eq. B1-B3 from \cite{Jodlowski:2023yne}, while the derivation of the equation below can be found in the included Mathematica notebook,
\begin{dmath}[labelprefix={eq:}]
  \sigma_{\gamma^{\prime} N \to a N} \simeq \frac{\alpha_{\mathrm{EM}} g_{a\gamma\gamma^{\prime}}^2 Z^2}{12} \left(\log \left(\frac{d}{1/a^2 - t_{\mathrm{max}}}\right)-2\right),
  \label{eq:Prim}
\end{dmath}
where $a=111 Z^{-1/3}/m_e$, $d=0.164\, \gev^2 A^{-2/3}$, $m_e$ is the electron mass, $Z$ ($A$) is the atomic number (weight) of a nucleus, and $t_{\mathrm{max}} \simeq -(m_a^4+m_{\gamma^{\prime}}^4)/(4E_1^2)$.

This formula differs (it is smaller) from the one for photophilic ALP only by a factor of $2/3$, which results from $2$ ($3$) polarization states for photon (dark photon).

At FASER$\nu$2, Primakoff production process takes place on tungsten (W), and the formula describing it has the following form: $\sigma^{\text{W}}_{\gamma^{\prime} N \to a N} \simeq \frac{48}{\mathrm{GeV}^2} \times \left(\frac{g_{a\gamma\gamma^{\prime}}}{1/\mathrm{GeV}}\right)^2$.

\paragraph{Electron scattering}
FASER$\nu$2 and FLArE detectors will be sensitive to DS states scattering with electrons, see \cite{Batell:2021blf} for an extensive discussion.

The corresponding probability for such scattering events is simply given by
\be
  p(E)_{\text{scat.}} = \frac{\Delta}{L_{\text{int}}},
  \label{eq:p_sec}
\ee
where $\Delta$ is the length of the FASER$\nu$2 or FLArE and $L_{\text{int}}$ denotes the interaction length of the scattering process, which is described by the following cross-section:
\begin{dmath}[labelprefix={eq:}]
  \sigma_{\gamma^{\prime} e^- \to a e^-} \simeq \frac{\alpha_{\mathrm{EM}} g_{a\gamma\gamma^{\prime}}^2}{12} \log \left( \frac{E_R^{max}}{E_R^{min}} \right),
  \label{eq:e_scat}
\end{dmath}
where $E_R\equiv E_{e^-}$ is the electron recoil energy, while the differential cross-section, $d\sigma/dE_R$, needed for angular cuts indicated in \cref{tab:experiments}, can be found in the Mathematica notebook.

%=============================================================================
\section{Results\label{sec:results}}
%=============================================================================

We present the sensitivity lines for the LLP signatures within DAP in past and future experiments listed in \cref{tab:experiments}.
In the upper rows, the characteristics of beam dump experiments are displayed, whereas the lower rows exhibit the far-forward LHC detectors.
We consider two versions of the FASER2 detector - an extension of current FASER detector or new one placed within FPF.
We indicated all their relevant properties needed for simulation of LLP signatures discribed in \cref{sec:LLPs_at_exp}.

Although, as indicated by similar formulas for decay widths and Primakoff cross-sections for both DAP and photophilic ALP, the main difference between the two portals is the fact that DAP is composed of two DS species and, depending on the mass hierarchy between them, each can serve as a LLP.

In \cref{sec:dark_photon_LLP} we discuss the case when dark photon is the LLP, which for massless dark axion was studied in \cite{deNiverville:2018hrc,deNiverville:2019xsx,Deniverville:2020rbv}, while in \cref{sec:dark_axion_LLP} we consider dark axion as the LLP.

%==========================================================================
\subsection{Dark photon as the LLP \label{sec:dark_photon_LLP}}
%==========================================================================

When $m_{\gamma^{\prime}} > m_a$, $\sim\,$sub-GeV dark photon is the LLP, and its decay width is described by \cref{eq:ctau_gprime}.
The signatures described in \cref{sec:LLPs_at_exp} were simulated in modified version of $\tt FORESEE$, and the results are shown in \cref{fig:results_dark_photon}.

We consider three mass ratios, $m_{a}/m_{\gamma^\prime}$, which are fixed as follows: $0$ (top), $0.5$ (middle), and $0.9$ (bottom).

In the first case, we checked that when the dark photon is produced only by the three-body pseudoscalar meson decays, we reproduce the results of \cite{deNiverville:2018hrc,deNiverville:2019xsx}.
Moreover, for the case of massless dark axion (and also in the opposite case of massless dark photon), we denote with the light-gray color the areas that are excluded by astrophysical and cosmological bounds obtained in \cite{Hook:2021ous}.

The richness of the DAP is indicated by the middle and bottom plots of \cref{fig:results_dark_photon}. 
They show that when the masses of the two DS particles are comparable, the LLP decay width is suppressed and, as a result, its lifetime is longer, resulting in shifting the significant reach of FASER2 and SHiP towards higher masses.
Note that in this scenario the existing bounds, especially from NuCal, are relaxed due to the high energy threshold on the single photon, which is more difficult to meet because of the compressed spectra.
On the other hand, FASER2 reach weakens only mildly because of the typical high energy $\sim O(100)'s\,\gev$ of the produced LLPs.

Another feature of DAP that distinguishes it from the photophilic ALP is that vector meson decays produce a pair of dark photon-dark axion, both of which can travel virtually undisturbed from the production point to FASER$\nu$2, which allows for the secondary LLP production by Primakoff-like upscattering\footnote{It also increases the number of events from electron scattering by a factor of 2.}. 
As a result, this production mode allows to cover part of the $d=\gamma c\tau \sim 1\,\m$ region of the parameter space, see dashed and dash-dotted lines in \cref{fig:results_dark_photon,fig:results_dark_axion}.
Note that the probability of LLP decay taking place inside the decay vessel in short-lived regime is exponentially suppressed, $p(E) \simeq e^{-L/d}$ for $d \ll L$, hence this region of the parameter space cannot be covered by a detector placed at a significant distance from the LLP production point.

Lastly, the electron scattering signature allows coverage of the low-mass regime and is complementary to the decays of the dark photons produced in both primary and secondary production processes.
It should be noted that the electron scattering limit is typically weaker than in the case of secondary production, mainly due to the lack of $Z^2$ enhancement, cf. \cref{eq:Prim} and \cref{eq:e_scat}.

%==========================================================================
\subsection{Dark axion as the LLP \label{sec:dark_axion_LLP}}
%==========================================================================

The results for the opposite mass hierarchy are shown in \cref{fig:results_dark_axion}.
The formulas for the LLP production channels are the same as for the case of the dark photon acting as the LLP, while the LLP lifetime is smaller by a factor of $3$; see \cref{eq:Gamma_two_body}.
As a result, the sensitivity lines are shifted towards smaller masses.
Moreover, the Primakoff and electron scattering cross-sections are also smaller by a factor of $3$, resulting in smaller reach.

We only show one benchmark corresponding to massless dark photon, while other mass scenarios are analogous to the middle and bottom rows of \cref{fig:results_dark_photon}.

%=============================================================================
\section{Conclusions\label{sec:conclusions}}
%=============================================================================

In this paper, we have studied the prospects of detecting the dark axion portal in the intensity frontier searches adapted to a diverse set of LLP signatures.
The main difference between DAP with negligible kinetic mixing and the photophilic ALP in the $\sim\,$sub-GeV mass regime is that the Primakoff conversion of an on-shell photon into an ALP is no longer possible, and the leading LLP production modes  are vector meson decays, yielding approximately an order of magnitude fewer LLPs.

The second difficulty in probing DAP is that the LLP decays semi-visibly, so its energy can only be deposited by a single photon\footnote{As \cref{fig:results_dark_photon,fig:results_dark_axion} show, three-body semi-visible LLP decays into a $e^+ e^-$ pair are suppressed by at least 2 orders of magnitude with respect to the leading two-body decays.}.
Such an experimental signature is more challenging than the usual two-photon ALP decay because of, among other things, the additional SM induced background.

On the other hand, future detectors like FASER2 and SHiP will be able to effectively probe such LLP decays, resulting in coverage of the parameter space similar to the photophilic ALP.

Moreover, secondary LLP production taking place just in front of the decay vessel will allow to cover part of the shorter LLP lifetime regime corresponding to $d\sim 1\,\m$.

Finally, scatterings of either of the DS species with electrons taking place inside FASER$\nu$2 or FLArE will allow to probe the low mass regime of the LLP, $m\lesssim 10\,\mev$. It is complementary to both the LLP displaced decays and missing energy searches, which are restricted to different LLP mass ranges.

%=============================================================================
\acknowledgments
%=============================================================================

This work was supported by the Institute for Basic Science under the project code, IBS-R018-D1.

% We thank Polish Ministry of Science and Higher Education, Foundation for Polish Science, and Polish National Agency for Academic Exchange for \textit{not} supporting this research.

\appendix

\section{Meson decays}

In this section, we present expressions for the leading dark axion and dark photon production modes, see \cref{fig:modes_production} for a comparison of their contributions.

% \section{Vector meson decays}
\subsection{Vector meson decays}
\label{app:prod}

We give the branching ratio of the two-body decays of vector mesons into $a{\text -}\gamma^{\prime}$, $V(p_0) \!\to\! \gamma^*(p_1+p_2) \!\to\! a(p_1) + \gamma^{\prime}(p_2)$, which are mediated by an off-shell photon, 
\be
  \frac{{\rm BR}_{V \rightarrow a \gamma^\prime}}{{\rm BR}_{V \rightarrow ee}} = \frac{ g_{a\gamma\gamma^\prime}^2 \left((-M^2+m_a^2+m_{\gamma^\prime}^2)^2-4 m_a^2 m_{\gamma^\prime}^2\right)^{3/2}}{32 \pi  \alpha_{\text{EM}} M \sqrt{M^2-4 m_e^2} \left(M^2+2 m_e^2\right)},
  \label{eq:br_vec}
\ee
where ${\rm BR}_{V\rightarrow e^+ e^-}$ is the branching ratio of the vector meson with mass $M$ decaying into the $e^+ e^-$ pair \cite{Workman:2022ynf}.

We note that in the $m_{\gamma^{\prime}} \to 0$ limit \cref{eq:br_vec} reduces to the result for photophilic ALP, see, \eg, Eq. 11 from \cite{Merlo:2019anv}.

%==============================================================================================
\subsection{Pseudoscalar and vector meson decays}
%==============================================================================================
\label{app:d2Br}

The subdominant production mode takes place by decays of pseudoscalar mesons into a photon and DS states mediated by an off-shell photon, $P(p_0) \!\to\!  \gamma(p_1)+ \gamma^*(p_2+p_3) \!\to\!  \gamma(p_1) + a(p_2) + \gamma^{\prime}(p_3)$.

We obtained the same averaged amplitude squared as \cite{deNiverville:2018hrc}, while below we give the resulting differential branching ratio in a form convenient for Monte Carlo simulation:
\begin{widetext}
\be
  \frac{d{\rm BR}_{P \rightarrow \gamma a \gamma^{\prime}}}{dq^2 d\cos\theta} = {\rm BR}_{P\rightarrow \gamma \gamma}  &\!\times \!\! \left[ \frac{g_{a\gamma\gamma^\prime}^2}{256 \pi ^2 m_P^6 q^6} \left(m_P^2-q^2\right)^3 (\cos(2\theta)+3) \left( (m_{\gamma^\prime}^2 + m_a^2 - q^2)^2 - 4 m_{\gamma^\prime}^2  m_a^2 \right)^{3/2}\!\right],
  \label{eq:br2dq2dcostheta}
\ee
\end{widetext}
where $m_P$ is the pseudoscalar meson mass, $q^2 \equiv (p_2+p_3)^2$, and $\theta$ is the angle between $a$($p_2$) and the off-shell photon in the meson rest frame; ${\rm BR}_{P\rightarrow \gamma \gamma}$ is the branching ratio of pseudoscalar meson decaying into two photons taken for PDG \cite{Workman:2022ynf}.

%==============================================================================================
\section{Three-body decays}
\label{app:three_body_decays}
%==============================================================================================

Below, we give formulas for the three-body decay widths of a dark photon and a dark ALP in the $m_{\gamma^{\prime}} \gg m_a, m_l$ and $m_a \gg m_{\gamma^{\prime}}, m_l$ limits, respectively; general form of the differential decay width can be found in the Mathematica notebook included in \href{https://github.com/krzysztofjjodlowski/Looking_forward_to_photon_coupled_LLPs}{\faGithub}.

\begin{widetext}
  \begin{dmath}[labelprefix={eq:}]
    \Gamma_{\gamma^{\prime}\to l^+ l^- a} = \frac{\alpha_{\mathrm{EM}} g_{a\gamma \gamma^{\prime}}^2}{576 \pi ^2 m_{\gamma}^3} \left(32 m_l^6 \coth^{-1}\left(\frac{m_{\gamma^{\prime}}}{\sqrt{m_{\gamma^{\prime}}^2-4 m_l^2}}\right) + m_{\gamma^{\prime}} \left(\sqrt{m_{\gamma^{\prime}}^2-4 m_l^2} \left(26 m_{\gamma^{\prime}}^2 m_l^2-7 m_{\gamma^{\prime}}^4+8 m_l^4\right)-4 m_{\gamma^{\prime}}^5 \log \left(\frac{2 m_l}{\sqrt{m_{\gamma^{\prime}}^2-4 m_l^2}+m_{\gamma^{\prime}}}\right)+12 m_{\gamma^{\prime}} m_l^4 \log \left(\frac{16 m_l^4 \left(m_{\gamma^{\prime}}-\sqrt{m_{\gamma^{\prime}}^2-4 m_l^2}\right)}{\left(\sqrt{m_{\gamma^{\prime}}^2-4 m_l^2}+m_{\gamma^{\prime}}\right)^5}\right)\right)\right),
    \label{eq:gprime_lla}
  \end{dmath}
\end{widetext}

\begin{widetext}
  \begin{dmath}[labelprefix={eq:}]
    \Gamma_{a\to l^+ l^- \gamma^{\prime}} = \frac{\alpha_{\mathrm{EM}} g_{a\gamma \gamma^{\prime}}^2}{192 \pi ^2 m_a^3} \left(32 m_l^6 \coth^{-1}\left(\frac{m_a}{\sqrt{m_a^2-4 m_l^2}}\right) + m_a \left(\sqrt{m_a^2-4 m_l^2} \left(26 m_a^2 m_l^2-7 m_a^4+8 m_l^4\right)-4 m_a^5 \log \left(\frac{2 m_l}{\sqrt{m_a^2-4 m_l^2}+m_a}\right)+12 m_a m_l^4 \log \left(\frac{16 m_l^4 \left(m_a-\sqrt{m_a^2-4 m_l^2}\right)}{\left(\sqrt{m_a^2-4 m_l^2}+m_a\right)^5}\right)\right)\right).
    \label{eq:a_llg}
  \end{dmath}
\end{widetext}

\bibliography{main_dark_alp}

\end{document}